\documentclass[aps,amsmath,amssymb,twocolumn]{revtex4-1}

\usepackage{graphicx}
\usepackage{dcolumn}
\usepackage{bm}
\usepackage{color}
\usepackage[normalem]{ulem}

\begin{document}

\title{Thermal discrete dipole approximation for near-field radiative heat transfer in many-body systems with arbitrary 
nonreciprocal bodies}

\author{E. Moncada-Villa$^{1}$}
\author{J.~C. Cuevas$^{2}$}

\affiliation{$^{1}$Escuela de F\'{\i}sica, Universidad Pedag\'ogica y Tecnol\'ogica de Colombia,
Avenida Central del Norte 39-115, Tunja, Colombia}

\affiliation{$^2$Departamento de F\'{\i}sica Te\'orica de la Materia Condensada
and Condensed Matter Physics Center (IFIMAC), Universidad Aut\'onoma de Madrid, E-28049 Madrid, Spain}

\date{\today}

\begin{abstract}
The theoretical study of many-body effects in the context of near-field radiative heat transfer (NFRHT) has already led to
the prediction of a plethora of thermal radiation phenomena. Special attention has been paid to nonreciprocal systems in
which the lack of the Lorentz reciprocity has been shown to give rise to unique physical effects. However, most of the theoretical 
work in this regard has been carried out with the help of approaches that consider either point-like particles or highly symmetric 
bodies (such as spheres), which are not easy to realize and explore experimentally. In this work we develop a many-body approach based 
on the thermal discrete dipole approximation (TDDA) that is able to describe the NFRHT between nonreciprocal objects of arbitrary size 
and shape. We illustrate the potential and the relevance of this approach with the analysis of two related phenomena, namely the 
existence of persistent thermal currents and the photon thermal Hall effect, in a system with several magneto-optical bodies. Our 
many-body TDDA approach paves the way for closing the gap between experiment and theory that is hindering the progress of the topic 
of NFRHT in many-body systems.     
\end{abstract}

\maketitle

\section{Introduction}

It is known that the Stefan-Boltzmann law sets an upper limit for the far field radiative thermal exchange between two bodies 
at different temperatures, which is valid for separations above the thermal wavelength ($\lambda_{\rm th}\sim 9.6$ $\mu$m at 
room temperature). However, if the bodies are brought in closer proximity (below $\lambda_{\rm th}$), the NFRHT contribution 
arising from the evanescent field of electromagnetic waves at the material surfaces may lead to largely overcome the blackbody 
limit. This near-field enhancement was first predicted by Polder and Van Hove back in 1971 \cite{Polder1971} within the framework 
of the fluctuational electrodynamics (FE) \cite{Rytov1989}, and in recent years it has been experimentally studied in a great 
variety of systems and using different types of materials  \cite{Kittel2005,Narayanaswamy2008,Hu2008,Rousseau2009,Shen2009,Shen2012,
Ottens2011,Kralik2012,Zwol2012a,Zwol2012b,Guha2012,Worbes2013,Shi2013,St-Gelais2014,Song2015b,Kim2015,Lim2015,St-Gelais2016,Song2016,
Bernardi2016,Cui2017,Kloppstech2017,Ghashami2018,Fiorino2018,DeSutter2019}. These experiments, in turn, have been crucial to 
firmly establish the basic NFRHT mechanisms in two-body systems, for recent reviews see Refs.~\cite{Song2015a,Cuevas2018}.

In this context, part of the attention of the thermal radiation community has shifted to the investigation of many-body
effects, a topic that is thus far largely dominated by the theory \cite{Biehs2021}. The NFRHT in many-body systems offers 
new exciting possibilities such as the development of functional devices, which are often the thermal radiation counterparts 
of electronic devices (diodes, transistors, etc.) \cite{Ben-Abdallah2017,Biehs2021}. On the other hand, many-body 
systems have also turned out to be ideal to unveil new physical phenomena that are absent in the standard two-body configuration. 
This is specially clear in the case of many-body systems involving nonreciprocal bodies, which are often based on magneto-optical 
(MO) objects whose near-field radiation properties can be largely tuned with external magnetic fields 
\cite{Moncada-Villa2015,Latella2017,Abraham-Ekeroth2018,Ott2019,Moncada-Villa2020,Moncada-Villa2021}. 
Thus, for instance, in 2016 Ben-Abdallah predicted the possibility of having a near-field thermal analog of the Hall effect 
in an arrangement of four MO particles placed in a constant magnetic field \cite{Ben-Abdallah2016}. Also in 2016, Zhou and Fan 
showed that a many-body system comprising MO nanoparticles can support a persistent directional heat current, without violating 
the second law of thermodynamics \cite{Zhu2016}. In fact, it has been shown that in certain systems these two striking phenomena 
are very much related and they should appear together \cite{Guo2019}.

The theoretical description of the NFRHT in nonreciprocal many-body systems has been restricted so far to either point-like 
particles \cite{Ben-Abdallah2016,Ott2020} or highly symmetric bodies \cite{Zhu2016,Guo2019,Zhu2018}. Thus, it would be 
desirable to develop new theoretical methods to describe the NFRHT in this type of systems with bodies of arbitrary size and 
shape. This is exactly the goal of this work. To be precise, we present here an approach to deal with the NFRHT in nonreciprocal
many-body systems made of objects of arbitrary size and shape which is based on an extension the TDDA method that has been 
successful describing the NFRHT of two-body systems \cite{Edalatpour2014,Edalatpour2015}, including those based on MO objects 
\cite{Abraham-Ekeroth2017,Abraham-Ekeroth2018}. The TDDA approach, which is very much related to existent approaches to describe
the radiative heat transfer in many-body systems of point dipoles \cite{Messina2013,Tervo2019}, is based on a 
natural extension of the discrete dipole approximation (DDA) that is widely used for describing the scattering and absorption 
of light by small particles \cite{Purcell1973,Draine1988,Draine1994,Yurkin2007}. We shall illustrate here the power of our many-body 
TDDA method by analyzing both the photon Hall effect and the existence of a persistent heat current in a system of MO particles 
beyond the standard point-dipole approximation. In particular, in the case of photon Hall effect, our theory amends the results
of the original work of Ref.~\cite{Ben-Abdallah2016} and we show that the appearance of this effect in particles of finite size 
does not require extremely high magnetic fields. 

The remainder of this paper is organized as follows. In Sec.~\ref{formalism} we present the many-body formalism based on the
TDDA approach that enables us to describe the NFRHT between nonreciprocal bodies of arbitrary size and shape. Section~\ref{app-hall} 
is devoted to the application of this formalism to the description of the photon thermal Hall effect and the appearance of a 
persistent heat current in a system comprised of four MO spherical particles. Finally, we summarize our main results in
Sec.~\ref{conclusions} and discuss possible future research lines.

\section{Radiative heat exchange in a system with arbitrary anisotropic bodies} \label{formalism}

\begin{figure}[t]
\includegraphics[width=\columnwidth,clip]{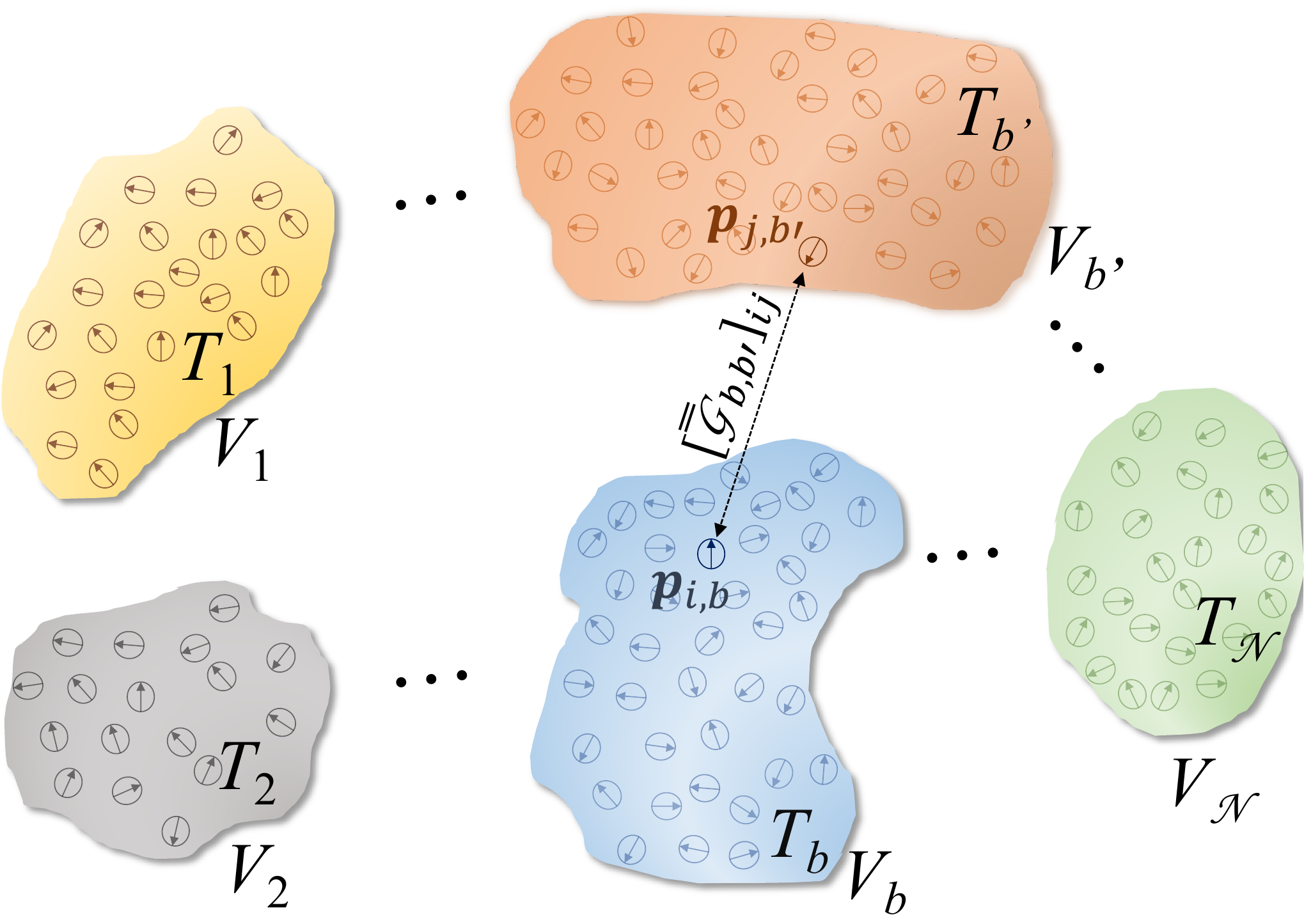}
\caption{Schematic representation of a generic many-body system consisting of $\mathcal N$ anisotropic bodies, each one at a 
temperature $T_b$ and with a total volume $V_b$, described by a collection of $N_{b}$ electrical point dipoles, $\mathbf{p}_{i,b}$, 
with a volume $V_{i,b}$ and dielectric permittivity tensor $\hat\varepsilon_{i,b}$.}
\label{fig-geom}
\end{figure}

The goal of this paper is to extend the TDDA approach to describe the near-field radiative heat transfer in many-body systems 
containing optically anisotropic bodies of arbitrary size and shape. We start by considering a system of $\mathcal N$ anisotropic 
bodies, as depicted in Fig.~\ref{fig-geom}, each one having its own temperature $T_b$ (the same throughout the body) and volume 
$V_b$. To describe this system we make use of the DDA \cite{Novotny}, in which each body $b$ is discretized in terms of a collection of 
$N_{b}$ electrical point dipoles of volume $V_{i,b}$, dielectric permittivity tensor $\hat\varepsilon_{i,b}$, and polarizability 
\begin{equation}\label{polariz}
	\hat\alpha_{i,b}= \left[ \frac{1}{V_{i,b}}(\hat L_{i,b} + [\hat\varepsilon_{i,b}-\hat 1 ]^{-1} ) - 
	i\frac{k_0^3}{6\pi}\hat 1 \right]^{-1},
\end{equation}
where $\hat L_j$ is the so-called depolarization tensor \cite{Lakhtakia1992,Yaghjian1980}, which for cubic volume elements is 
diagonal and equal to $(1/3)\hat 1$. The computation of the radiative heat transfer between the $\mathcal N$ bodies starts with 
the calculation of the statistical average of the power dissipated in a particular body $b$ due to the emission of the 
$\mathcal N -1$ remaining bodies \cite{Abraham-Ekeroth2017}, that is
\begin{equation}\label{power}
	\mathcal P_b = \int_{V_b} \langle \bar{\mathbf J}_b(t) \cdot \bar{\mathbf E}_b(t) \rangle d \mathbf r =  
	\langle \frac{d\bar{\mathbf P}_b(t)}{dt}\cdot \bar{\mathbf E}_b(t) \rangle,
\end{equation}
where $\bar{\mathbf P}_b = \left(\mathbf p_{1}, \mathbf p_{2},...,\mathbf p_{N_{b}}\right)^{{\rm T}}$ is a column supervector of 
dimension $3N_b\times 1$, whose components are the $N_b$ electrical point dipolar moments in body $b$, the $3N_b\times 1$ 
supervector $\bar{\mathbf E}_b = \left(\mathbf E_{1}, \mathbf E_{2},...,\mathbf E_{N_{b}}\right)^{{\rm T}}$ contains all the 
internal electric fields in each of the elementary volumens, and $\bar{\mathbf J}_b(t)$ contains all the local current densities 
related to the each electrical point dipole. For time harmonic fields, the absorbed power of Eq.~\eqref{power} can be written as
\begin{equation}\label{power-b}
	\mathcal P_b = 2\int_0 ^\infty \frac{d\omega}{2\pi} \omega \int_0 ^\infty \frac{d\omega^\prime}{2\pi}
	{\rm Im Tr}\{ \langle \bar{\mathbf E}_b^\dagger(\omega^\prime)\bar{\mathbf P}_b(\omega)\rangle e^{-i(\omega-\omega^\prime)t}\},
\end{equation}
where, for mathematical convenience, we use ${\rm Tr}\{\bar{\mathbf E}_b^\dagger(\omega^\prime)\bar{\mathbf P}_b(\omega)\}$ instead 
of $\bar{\mathbf P}_b(\omega) \cdot \bar{\mathbf E}_b^*(\omega^\prime)$. Each internal dipolar field contained in the vector 
$\bar{\mathbf E}_b$ appearing in Eq.~\eqref{power-b} have contributions from all dipolar moments in the system. Within the DDA approach, 
the self-consistent volume integral equations relating the internal fields and the dipolar moments for all the $\mathcal N$ objects 
adopt the following matrix form \cite{Novotny,Abraham-Ekeroth2017}
\begin{equation}\label{E-p}
	\bar{\bar{\mathbf E}} = \frac{k_0^2}{\varepsilon_0}\bar{\bar{  \mathcal G }}\hspace{0.1cm}\bar{\bar{\mathbf P}}, 
\end{equation}
where $\bar{\bar{\mathbf E}} = \left(\bar{\mathbf E}_1,\bar{\mathbf E}_2,...,\bar{\mathbf E}_{\mathcal N} \right)^{{\rm T}}$, 
$\bar{\bar{\mathbf P}} = \left(\bar{\mathbf P}_1,\bar{\mathbf P}_2,...,\bar{\mathbf P}_{\mathcal N} \right)^{{\rm T}}$ and 
\begin{eqnarray}\label{E-p-0} 
	\bar{\mathbf P}_b&=&\varepsilon_0 {\rm diag}\left(V_{1,b} [ \hat\varepsilon_{1,b}-\hat 1 ] ,  ...,V_{N_b,b} 
	[ \hat\varepsilon_{N_b,b}-\hat 1 ] \right) \bar{\mathbf E}_b.
\end{eqnarray}

\noindent
Notice that $\bar{\bar{\mathbf E}}$ and $\bar{\bar{\mathbf P}}$ are $(\sum_{b=1}^{\mathcal N} 3N_b)\times 1$ supervectors 
containing, respectively, the internal electric fields and the dipolar moments of all the electric point dipoles in the 
$\mathcal N$-body system. In the matrix system of Eq.~\eqref{E-p}, $k_0=\omega/c$  and $\varepsilon_0$ are, respectively, 
the vacuum wave vector and electric permittivity. Moreover, the quantity $\bar{\bar{  \mathcal G}}$ denotes the dyadic Green's 
function matrix whose elements, connecting the $i$-th dipolar moment inside the body $b$ with the $j$-th dipolar moment of 
body $b^\prime$ (see Fig.~\ref{fig-geom}), and its elements are given by
\begin{eqnarray}
\label{green-1}
[\bar{ \bar{ \mathcal G}}_{b,b^\prime}]_{ij} & = & \left\{ \begin{array}{ll}
-\frac{\hat L_{i,b}}{V_{i,b}k_0^2} + \frac{ik_0}{6\pi}\hat 1 , & \;\; {\rm if} \; \; i=j \;\; \textrm{and} \;\; b = b^{\prime} \\
\left[ {\hat G}_{b,b^\prime} \right]_{ij} , & \;\; {\rm otherwise} 
\end{array} \right.
\end{eqnarray}
where
\begin{eqnarray}\label{green}
 [ {\hat G}_{b,b^\prime} ]_{ij} & = & \frac{e^{ik_0R}}{4\pi R} \left[ \left(1+\frac{ik_0R-1}{(k_0R)^2}\right)\hat 1 \right. 
 \\\nonumber && + \left. \left( \frac{3-3ik_0R-(k_0R)^2}{(k_0R)^2}\right) \frac{\mathbf R\otimes \mathbf R}{R^2 }\right], 
 \end{eqnarray}
and $\mathbf R=\mathbf r_{i,b} - \mathbf r_{j,b^\prime} $. In problems like ours, where we are interested in the radiative heat 
exchange, the electrical dipoles have two basic contributions, one is a fluctuating one related to the thermal emission and the 
second one is an induced part related to the interactions between dipoles. Therefore, the total dipole moments can be written 
as \cite{Messina2013}
\begin{equation}\label{p-tot}
	\bar{\bar{\mathbf P}} = \bar{\bar{\mathbf P}}^{{\rm (ind)}} + \bar{\bar{\mathbf P}}^{{\rm (fl)}} .
\end{equation}
Thus, in order to compute the thermal average of ${\rm Tr}\{\bar{\mathbf E}_b^\dagger(\omega^\prime)\bar{\mathbf P}_b(\omega)\}$
in Eq.~\eqref{power-b}, our task now is to express the quantities $\bar{\bar{\mathbf E}}_b$  and $\bar{\bar{\mathbf P}}_b$ in terms 
of the fluctuating parts of the dipolar moments, whose properties are known (via the fluctuation-dissipation theorem). For this
purpose, let us start by pointing out that the induced part of the local dipole moment is caused by all the local fields, but its 
own, that is
\begin{equation}\label{p-ind}
	\bar{\bar{\mathbf P}}^{{\rm (ind)}} =\varepsilon_0 \bar{\bar{\alpha}} \bar{\bar{\mathbf E}}_{\rm exc},
\end{equation}
where $ \bar{\bar{\alpha}}={\rm diag}( \bar{\alpha}_1, \bar{\alpha}_2,...,  \bar{\alpha}_{\mathcal N})$ is a block-diagonal matrix 
with dimensions $(\sum_{b=1}^{\mathcal N} 3N_b)\times (\sum_{b=1}^{\mathcal N} 3N_b)$, whose elements are the $3N_b\times 3N_b$ block 
diagonal matrices $ \bar{\alpha}_b={\rm diag}( \hat{\alpha}_{1,b},\hat{\alpha}_{2,b},...,\hat{\alpha}_{N_b,b})$, and $\hat\alpha_{i,b}$ 
are the electrical polarizabilities given by Eq.~\eqref{polariz}. The exciting field, $\bar{\bar{\mathbf E}}_{\rm exc}$, can be obtained 
after the subtraction of the self-interaction terms from the total internal field $\bar{\bar{\mathbf E}}$ as follows
\begin{equation}\label{E-exc}
  \bar{\bar{\mathbf E}}_{\rm exc}= \frac{k_0^2}{\varepsilon_0} \Delta\bar{\bar{ \mathcal G}}\hspace{0.1cm}\bar{\bar{\mathbf P}},
\end{equation}
where $\Delta\bar{\bar{ \mathcal G}}=\bar{\bar{ \mathcal G}}-{\rm diag}(\bar{\bar{ \mathcal G}})$. Insertion of this exciting field 
into the Eq.~\eqref{p-ind} leads to the following expression for the induced dipolar moment 
\begin{equation}\label{p-ind-2}
	\bar{\bar{\mathbf P}}^{{\rm (ind)}}=k_0^2 \bar{\bar{\alpha}}\Delta\bar{\bar{ \mathcal G}}\hspace{0.1cm}\bar{\bar{\mathbf P}},
\end{equation}
from which, with the aid of Eq.~\eqref{p-tot}, one obtains
\begin{equation}\label{p-ind-fl}
	\bar{\bar{\mathbf P}} =\bar{\bar D}^{-1}\bar{\bar{\mathbf P}}^{{\rm (fl)}},
\end{equation}
where $\bar{\bar D}=\hat 1 - k_0^2\bar{\bar\alpha}\Delta\bar{\bar{ \mathcal G}}$. From this, one arrives at the desired relation 
between the dipolar moments in the body $b$ and the fluctuating part of all the dipoles of the $\mathcal N$-body system
\begin{eqnarray}\label{p-ind-fl-1}
		{\bar{\mathbf P}} _b&=&\sum_{b^\prime} [\bar{\bar D}^{-1} ]_{bb^\prime}{\bar{\mathbf P}}^{{\rm (fl)}}_{b^\prime}.
\end{eqnarray}

At this point, we still have to calculate the internal fields in the body $b$, $\bar{\mathbf E}_b$, in terms of the fluctuating 
part of all the dipoles in the $\mathcal N$-body system. In what follows, we present two different, but equivalent ways in which
this calculation can be done. 

\subsection{Method 1}

As a first approach, we follow a line of reasoning similar to that used by Messina \textit{et al.} in Ref.~\cite{Messina2013} for
point-like particles. In this case, we start by inserting the relation of Eq.~\eqref{p-ind-fl} in Eq.~\eqref{E-p}, to obtain
\begin{equation}\label{super-E2}
	\bar{\bar{\mathbf E}}  = \frac{1}{\varepsilon_0}\bar{\bar{\mathcal C}} \hspace{0.1cm}\bar{\mathbf P}^{{\rm (fl)}},
\end{equation}
from which
\begin{eqnarray} \label{p-ind-fl-2}
		\bar{\mathbf E} _b^\dagger & = & \frac{1}{\varepsilon_0}\sum_{b^\prime} \bar{\mathbf p}^{{\rm (fl)}\dagger}_{b^\prime}[\bar{\bar{\mathcal C}} ^\dagger]_{b^\prime b}, 
\end{eqnarray}
with $\bar{\bar{\mathcal C}}= k_0^2 \bar{\bar{\mathcal G}} \hspace{0.1cm}\bar{\bar D}^{-1}$. The previous relation, together with 
Eq.~\eqref{p-ind-fl-1}, allows us to write the net power in Eq.~\eqref{power-b} as
\begin{eqnarray}\label{power-b-2}\nonumber
	\mathcal P_b & = & \frac{2}{\varepsilon_0} \sum_{b^\prime,b^{\prime\prime}} \int_0 ^\infty \frac{d\omega}{2\pi}
	\omega\int_0 ^\infty \frac{d\omega^\prime}{2\pi} {\rm Im Tr}\{ [\bar{\bar{\mathcal C}} ^\dagger]_{b^\prime b} 
	[\bar{\bar D}^{-1}]_{bb^{\prime\prime}}   \\ & & \hspace{1.3cm} \times
	 \langle  \bar{\mathbf p}^{{\rm (fl)}}_{b^{\prime\prime}} (\omega^\prime)  
	 \bar{\mathbf p}^{{\rm (fl)}\dagger}_{b^\prime}  (\omega)\rangle e^{-i(\omega-\omega^\prime)t}\}.
\end{eqnarray}
On the other hand, the statistical average appearing in the preceding equation is given by the fluctuation dissipation theorem
\cite{Landau1980,Keldysh1994,Joulain2005}
\begin{equation}\label{fdt}
\langle  \bar{\mathbf p}^{{\rm (fl)}}_{b^{\prime\prime}} (\omega^\prime) \bar{\mathbf p}^{{\rm (fl)}\dagger}_{b^\prime} (\omega) \rangle 
= 4\pi\hbar \varepsilon_0\delta_{b^{\prime\prime}b^\prime} \delta(\omega-\omega^\prime) n_{\rm B}(\omega,T_{b^\prime})
\bar \chi _{b^\prime},
\end{equation}
where $n_{\rm B}(\omega,T_{b^\prime})=1/(e^{\hbar\omega/k_{\rm B} T_{b^{\prime}}} -1 )$ is the Bose function, 
$\bar \chi _{b^\prime}={\rm diag}\left( \hat \chi _{1,b^\prime},\hat \chi _{2,b^\prime},...,\hat \chi _{N_{b^\prime},b^\prime} \right)$, 
and
\begin{equation}\label{susc}
	\hat \chi_{j,b} = \frac{1}{2i}( \hat \alpha _{j,b} -\hat \alpha _{j,b} ^\dagger) - 
	\frac{k_0^3}{6\pi} \hat \alpha_{j,b} ^\dagger\hat \alpha_{j,b}.
\end{equation}
Introducing Eq.~\eqref{fdt} into Eq.~\eqref{power-b-2} we can simplify the expression for the net power received by the body 
$b$ as follows
\begin{eqnarray}\label{power-b-3} \nonumber
	\mathcal P_b & = &4\sum_{b^\prime}\int_0 ^\infty \frac{d\omega}{2\pi} \hbar\omega n_{\rm B}(\omega,T_{b^\prime}){\rm Im Tr}\{ 
	[\bar{\bar{\mathcal C}} ^\dagger]_{b^\prime b}  [\bar{\bar D}^{-1}]_{bb^{\prime}} 
	\bar \chi _{b^\prime} \} . \\
\end{eqnarray}

In thermal equilibrium, when $T_b = T$ for all $b$, $\mathcal P_b$ must vanish, which implies that the following relation must
hold
\begin{eqnarray}\label{constraint}\nonumber
	{\rm Im Tr}\{ 
	[\bar{\bar{\mathcal C}} ^\dagger]_{b b}  [\bar{\bar D}^{-1}]_{bb} 
	\bar \chi _{b} \}=-\sum_{b^\prime\neq b}{\rm Im Tr}\{ 
	[\bar{\bar{\mathcal C}} ^\dagger]_{b^\prime b}  [\bar{\bar D}^{-1}]_{bb^{\prime}} 
	\bar \chi _{b^\prime} \},\\
\end{eqnarray}
which allows us to write the net power in Eq.~\eqref{power-b-3} as
\begin{eqnarray}\label{power-b-4}
	\mathcal P_b &=&\sum_{b^\prime\neq b} \mathcal P_{bb^\prime}  \\ \nonumber&=& \sum_{b^\prime\neq b}\int_0 ^\infty \frac{d\omega}{2\pi} [ \Theta(\omega,T_{b^\prime}) - \Theta(\omega,T_{b}) ]\tau_{bb^\prime},
\end{eqnarray}
where $\Theta(\omega,T_{b}) = \hbar\omega n_{\rm B}(\omega,T_{b})$ and $\tau_{bb^\prime}$ is the transmission probability between 
bodies $b$ and $b^\prime$, which is given by
\begin{equation}\label{trans-1}
	\tau_{bb^\prime}= 4 {\rm Im Tr}\{ [\bar{\bar{\mathcal C}} ^\dagger]_{b^\prime b}  [\bar{\bar D}^{-1}]_{bb^{\prime}} 
	\bar \chi _{b^\prime} \},
\end{equation}
Notice that transmission probability of Eq.~\eqref{trans-1} is different from that reported in Ref.~\cite{Ben-Abdallah2016}, 
which depends upon the imaginary part of the susceptibility instead on the full susceptibility function. As we will illustrate 
in Sec.~\ref{app-hall}, this difference leads to notable differences in the numerical results.

\subsection{Method 2}

Let us detail now an alternative way to compute the heat exchanges between the different bodies. In this case we use the relation 
of Eq.~\eqref{E-p-0}, instead of Eq.~\eqref{E-p}, to write
\begin{eqnarray}\label{E-p-2} 
	\bar{\mathbf E}_b = \bar \beta_b \bar{\mathbf P}_b,
\end{eqnarray}
with
\begin{equation}
\bar \beta_b = \frac{1}{\varepsilon_0} {\rm diag}\left( \frac{1}{V_{1,b}} [ \hat\varepsilon_{1,b}-\hat 1 ]^{-1} ,  ...,
\frac{1}{V_{N_b,b}} [ \hat\varepsilon_{N_b,b}-\hat 1 ]^{-1}  \right) .
\end{equation}
This relation, together with Eq.~\eqref{p-ind-fl-1} for the fluctuating part of the dipolar moment, 
allow us to write the dissipated power in body $b$, see Eq.~\eqref{power-b}, as 
\begin{eqnarray}\label{power-b-5}\nonumber
	\mathcal P_b & = & 2 \sum_{b^\prime b^{\prime\prime}} \int_0 ^\infty \frac{d\omega}{2\pi} \omega 
	\int_0 ^\infty \frac{d\omega^\prime}{2\pi} {\rm Im Tr}\{ [\bar{\bar D}^{-1} ]_{b^\prime b} ^\dagger \bar{\beta}_b^\dagger  
	[\bar{\bar D}^{-1} ]_{bb^{\prime\prime}}    \\ & & 
	\times \langle \bar{\mathbf P}_{b^{\prime\prime}}^{\rm (fl)}(\omega^\prime)  \bar{\mathbf P}^{{\rm (fl)}\dagger}_{b^\prime}(\omega)
	\rangle e^{-i(\omega-\omega^\prime)t}\}.
\end{eqnarray}
Using again the fluctuation dissipation theorem, see Eq.~\eqref{fdt}, we can write the previous relation as 
\begin{eqnarray}\label{power-b-6}
	\mathcal P_b & = & 4\varepsilon_0 \sum_{b^\prime} \int_0 ^\infty \frac{d\omega}{2\pi} \Theta(\omega,T_{b^\prime})  
	{\rm Im Tr}\{ [\bar{\bar D}^{-1} ]_{b^\prime b} ^\dagger \bar{\beta}_b^\dagger  \\\nonumber && \hspace{5cm} 
	\times [\bar{\bar D}^{-1} ]_{bb^{\prime}} \bar\chi_{b^\prime}\}\\ \nonumber
	& = & 4 \sum_{b^\prime} \int_0 ^\infty \frac{d\omega}{2\pi} \Theta(\omega,T_{b^\prime})  {\rm Tr}\{ [\bar{\bar D}^{-1} ]_{b^\prime b}
	 ^\dagger \frac{\varepsilon_0}{2i}(\bar{\beta}_b^\dagger-\bar{\beta}_b)  \\ \nonumber & & \hspace{5cm}
	 \times [\bar{\bar D}^{-1} ]_{bb^{\prime}} \bar\chi_{b^\prime}\}.
\end{eqnarray}
Additionally, it is straightforward to show that $\frac{\varepsilon_0}{2i} (\bar{\beta}_b^\dagger-\bar{\beta}_b) = 
(\bar\alpha_b^\dagger)^{-1}\bar\chi_b\bar\alpha_b^{-1}$, and therefore the net power absorbed by the body $b$
can be written as in Eq.~\eqref{power-b-4}, but with the following expression for the transmission probability 
\begin{eqnarray}\label{trans-2}
	\tau_{bb^\prime} & = & 4 {\rm Tr}\{ \bar{\alpha}_b^{-1} [\bar{\bar D}^{-1}]_{bb^{\prime}}\bar{\chi}_{b^\prime} 
	[\bar{\bar D}^{-1}]_{b^{\prime}b}(\bar{\alpha}_b^{\dagger })^{-1} \bar{\chi} _{b^\prime} \}.
\end{eqnarray}

As will be shown in the next section, the numerical results obtained with this relation are identical to those obtained with 
Eq.~\eqref{trans-1}. Furthermore, we would like to point out that, for a system consisting of only two arbitrary optically
anisotropic bodies, the previous relation reduces to 
\begin{eqnarray}\label{power-b-two-part}
	\tau_{bb^\prime} & = & 4 k_0^4 {\rm Tr} \{ \bar C_{21} \bar\chi_1 \bar C_{21}^\dagger\bar \chi_2 \},
\end{eqnarray}
with
\begin{eqnarray}\label{power-b-two-part-2}
	\bar{C}_{21} & = & \bar{\mathcal D}_{22}\Delta\bar {\mathcal D}_{22}
	[ \hat 1 - k_0^2 \bar \alpha _1 \Delta \bar{\bar{\mathcal G}}_{11}]^{-1},\\
	\bar {\mathcal D}_{22} & = & [\hat 1-k_0^2\Delta \bar{\bar{\mathcal G}}_{22}\bar\alpha_2 \\\nonumber & & 
	-k_0^4\Delta \bar{\bar{\mathcal G}}_{21}\bar\alpha_1[1-k_0^2\Delta\bar{\bar{\mathcal G}}_{11}\bar{\alpha}_1]^{-1}
	\Delta\bar{\bar {\mathcal G}}_{12}\bar{\alpha}_2]^{-1},
\end{eqnarray}
and $\bar {\mathcal G}_{bb^\prime}$ defined in Eq.~\eqref{green}. This expression agrees with the result of Eq.~(57)
reported in Ref.~\cite{Abraham-Ekeroth2017} for the two-body case.

\begin{figure}[t]
\includegraphics[width=\columnwidth,clip]{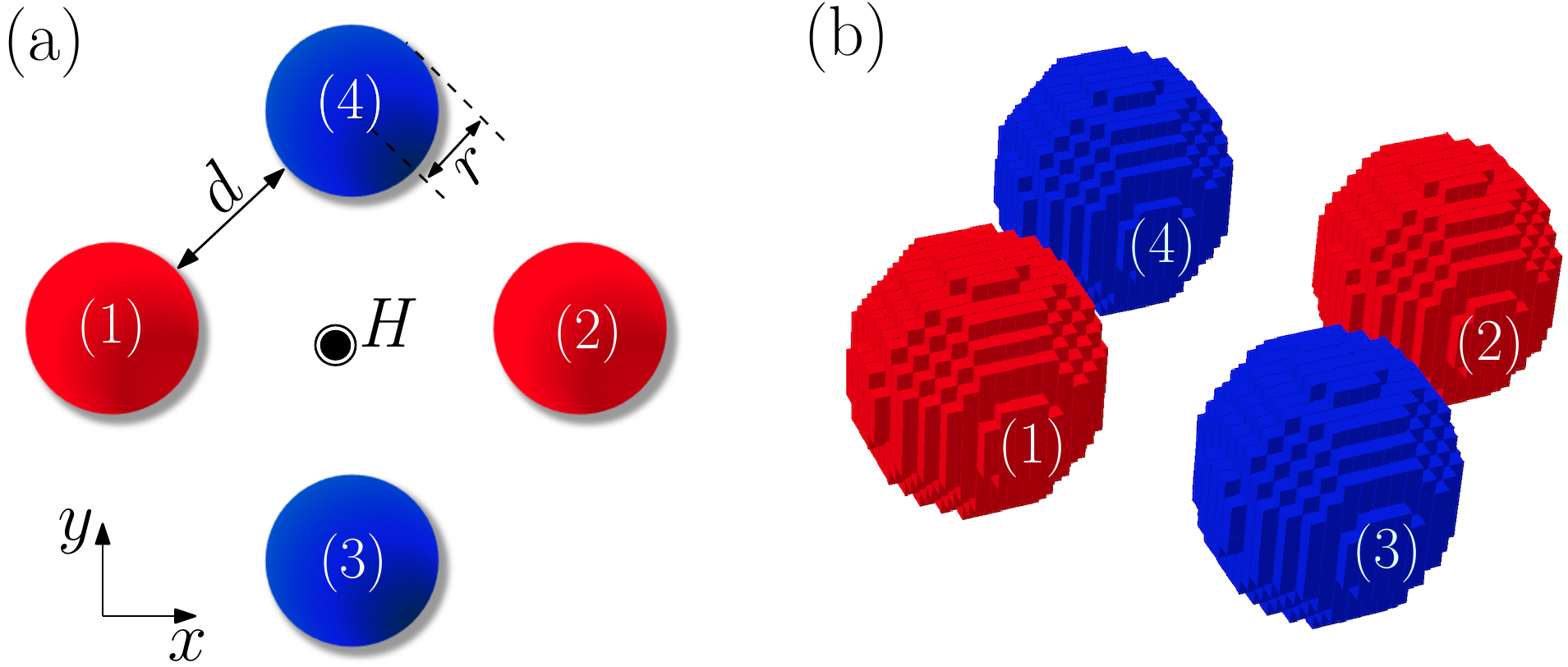}
\caption{(a) Schematic representation of the system under study, consisting of four magneto-optical spheres of radius $r$, 
placed in the vertices of a square in the $xy$ plane, and under the action of a static magnetic field $H$ along the positive 
$z$ direction. (b) Discretization each sphere as a collection of $N_{b}$ of cubic point dipoles used to employe the approach 
of the Sec.~\ref{formalism}.}
\label{fig-system}
\end{figure}

\section{Application to the analysis of the photon thermal Hall effect and the persistent heat current} \label{app-hall}

\begin{figure}[t]
\includegraphics[width=\columnwidth,clip]{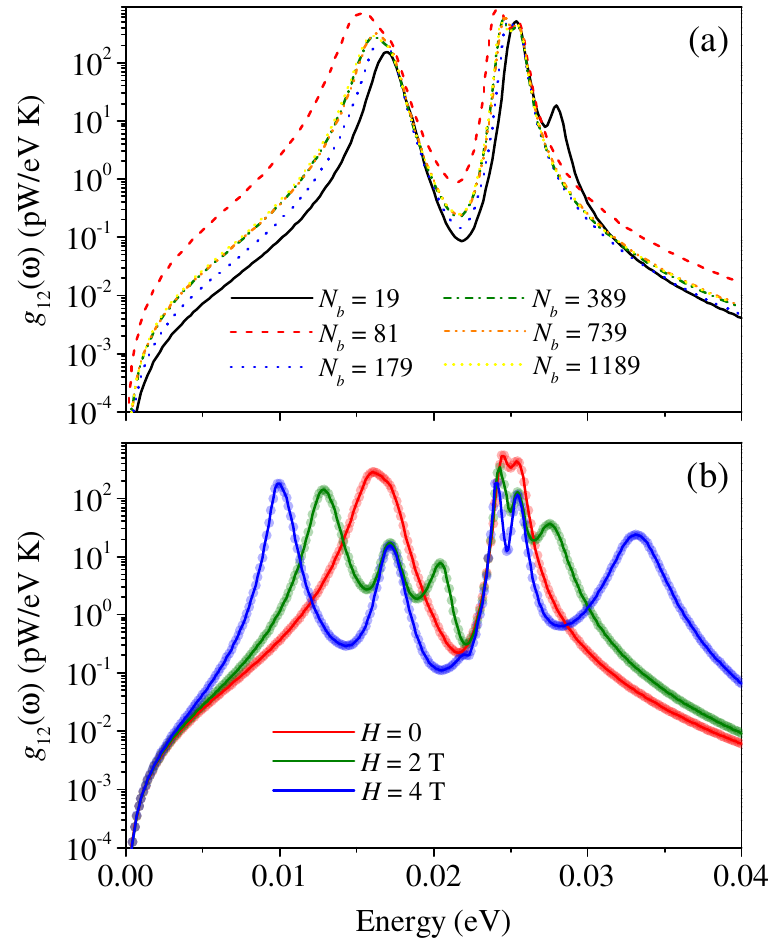}
\caption{(a) Zero-field spectral conductance as a function of the energy for spheres of a radius $r=100$ nm and separated 
by a vacuum gap of $d=20$ nm, for different values of the number of dipoles (see Fig.~\ref{fig-system}). (b) Spectral 
conductance for the same geometrical parameters in panel (a), with a fixed number of point dipoles ($N_b=389$). Continuous 
lines correspond to transmission probability calculated with Eq.~\eqref{trans-1}, whereas the circles correspond to the 
results obtained with Eq.~\eqref{trans-2}.}
\label{fig-conv}
\end{figure}

In this section we illustrate the use of the TDDA approach described in Sec.~\ref{formalism} with the analysis of 
two related many-body phenomena that occur with MO objects, namely the photon Hall effect and the possibility of having a persistent
heat current. For this purpose, and following Ref.~\cite{Ben-Abdallah2016}, we consider the system depicted in Fig.~\ref{fig-conv}(a) 
that comprises four identical and spherical particles made of the $n$-doped semiconductor indium antimonide (InSb). These particles
are eventually under the action of an external magnetic field along the positive $z$-direction, which makes these particles
nonreciprocal (see below). In the remainder of this paper, we shall assume these particles to be placed at the vertices 
of a square with a separation $d=20$ nm and assume that they have a radius of $r = 100$ nm. As demonstrated in 
Ref.~\cite{Ben-Abdallah2016}, if one creates a temperature difference $\Delta T = T_1 - T_2$ between particles 1 and 2, see 
Fig.~\ref{fig-conv}(a), the presence of an external magnetic field will cause particles 3 and 4, whose temperatures are left free, 
to be different ($T_3 \neq T_4$), which is the near-field thermal analog of the electronic Hall effect. The magnitude of this Hall 
effect can be evaluated using the relative Hall temperature difference
\begin{equation}
R = \frac{T_3 - T_4}{T_1 - T_2} .	
\end{equation}
In the linear response regime, in which $\Delta T$ is infinitesimally small, it can be shown that $R$ can be expressed as
\cite{Ben-Abdallah2016}
\begin{equation}\label{hall}
	R = \frac{g_{13} g_{24} -g_{14} g_{23}} {\sum_{b\neq 3} g_{b3} \sum_{b\neq 4} g_{b4} - g_{34}g_{43}},
\end{equation}
where $g_{bb^\prime}$ is the linear conductance between bodies $b$ and $b^\prime$ at an equilibrium temperature 
$T$ given by
\begin{equation} \label{cond}
	g_{bb^{\prime}} = \frac{\partial \mathcal P_{bb^\prime}}{\partial T} = 
	\int_0^\infty  \frac{d\omega}{2\pi}\frac{\partial\Theta(\omega,T)}{\partial T} \tau_{bb^\prime}(\omega).
\end{equation}
In the previous expression, we identify 
\begin{equation} \label{spec-cond}
	g_{bb^\prime}(\omega)=\frac{1}{2\pi}\frac{\partial\Theta(\omega,T)}{\partial T} \tau_{bb^\prime}(\omega)
\end{equation}
as the thermal conductance per frequency interval or spectral thermal conductance.

To compute the relative Hall thermal temperature difference we need to describe the optical properties of these MO
particles in the presence of a magnetic field. Following Ref.~\cite{Palik1976}, we assume that the InSb particles
become nonreciprocal and their field-dependent dielectric tensor is given by 
\begin{equation}
\label{perm-tensor-theta}
\hat \epsilon = 
\left( \begin{array}{ccc} 
\epsilon_1  &  -i\epsilon_2  &  0 \\
i\epsilon_2 &   \epsilon_1   &  0 \\
           0      &0 &  \epsilon_3 
\end{array} \right) ,
\end{equation}
where
\begin{eqnarray}
\epsilon_1(H) & = & \epsilon_{\infty} \left( 1 + \frac{\omega^2_L - \omega^2_T}{\omega^2_T - 
\omega^2 - i \Gamma \omega} + \frac{\omega^2_p (\omega + i \gamma)}{\omega [\omega^2_c -
(\omega + i \gamma)^2]} \right) , \nonumber \\
\label{eq-epsilons}
\epsilon_2(H) & = & \frac{\epsilon_{\infty} \omega^2_p \omega_c}{\omega [(\omega + i \gamma)^2 -
\omega^2_c]} , \\
\epsilon_3 & = & \epsilon_{\infty} \left( 1 + \frac{\omega^2_L - \omega^2_T}{\omega^2_T -
\omega^2 - i \Gamma \omega} - \frac{\omega^2_p}{\omega (\omega + i \gamma)} \right) . \nonumber
\end{eqnarray}
Here, $\epsilon_{\infty}$ is the high-frequency dielectric constant, $\omega_L$ ($\omega_T$) is  the longitudinal 
(transverse) optical phonon frequency, $\omega^2_p =ne^2/(m^{\ast} \epsilon_0  \epsilon_{\infty})$ is the plasma frequency of free 
carriers of density $n$ and effective mass $m^{\ast}$, $\Gamma$ ($\gamma$) is the phonon (free-carrier) damping constant, and 
$\omega_c =  eH/m^{\ast}$ is the cyclotron frequency, which depends on the intensity of the external magnetic  field. For the
moment, we assume the following room-temperature parameter values taken from Ref.~\cite{Palik1976}: $\epsilon_{\infty}  = 15.7$, 
$\omega_L = 3.62 \times 10^{13}$ rad/s, $\omega_T = 3.39\times 10^{13}$ rad/s, $\Gamma = 5.65 \times 10^{11}$ rad/s, 
$\gamma = 3.39 \times 10^{12}$ rad/s, $n = 1.07 \times 10^{17}$  cm$^{-3}$, $m^{\ast}/m = 0.022$, $\omega_p = 3.14 \times 10^{13}$ rad/s.

We start the analysis of our results by exploring the convergence of our method. In Fig.~\ref{fig-conv}(a), we show the spectral heat
conductance $g_{12}(\omega)$ from body 1 to 2, see Fig.~\ref{fig-system}, calculated with the expression of Eq.~\eqref{trans-1} 
transmission probability Eq.~\eqref{trans-1}. In this case there is no external magnetic field and the different curves corresponds
to different numbers of point dipoles used to model each spherical particle. Notice the fast convergence of our method, which requires 
only 389 electrical point dipoles to reach convergence (within a relative error of \%1). This is much more efficient than the method
presented in Ref.~\cite{Abraham-Ekeroth2017} for the two body problem, which requires around 2553 point dipoles for a similar converge. 
Such a difference is due to the fact that in the current method one does not do as many matrix operations as in 
Ref.~\cite{Abraham-Ekeroth2017}, as can be seen in Eqs.~\eqref{power-b-two-part} and \eqref{power-b-two-part-2}, which avoids 
error propagation. 

Once the convergence with the number of the point dipoles has been tested, we fix such a number to 389 and proceed to compare the 
results of the spectral heat transfer obtained with both transmission probabilities, see Eqs.~\eqref{trans-1} and \eqref{trans-2}. 
The results displayed in Fig.~\ref{fig-conv}(b) show that, within the numerical accuracy, both methods give the same result for 
different values of the magnetic field. It is worth stressing that we have found this level of agreement for all the configurations 
investigated and therefore, we conclude that these two methods are equivalent. With this in mind, all the results shown in what 
follows were obtained with transmission probability given by Eq.~\eqref{trans-1}.

\begin{figure}[t]
\includegraphics[width=\columnwidth,clip]{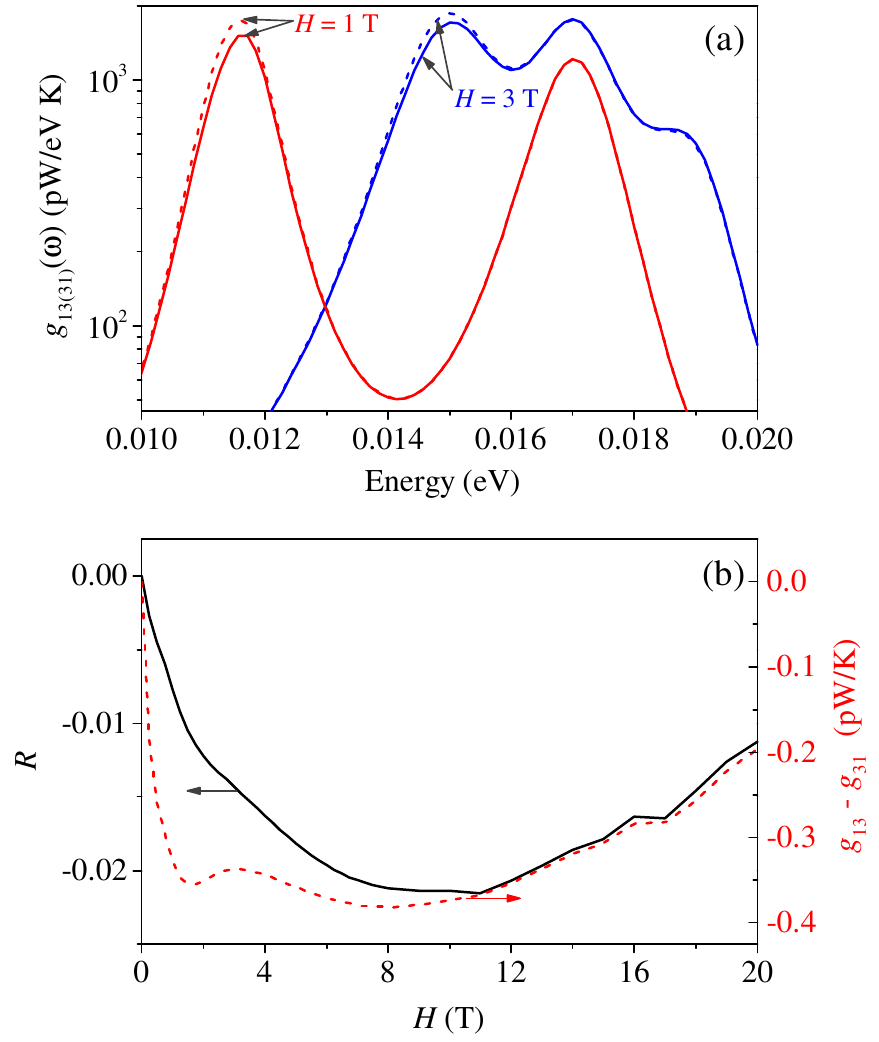}
\caption{(a) Spectral conductance $g_{13}(\omega)$ from the sphere 1 to 3 (solid lines), and from the 3 to 1, $g_{31}(\omega)$ 
(dashed lines), as a function of the photon energy, for two different values of applied magnetic field. (b) Hall coefficient, 
Eq.~\eqref{hall}, and persistent heat current $g_{13}-g_{31}$, Eq.~\eqref{cond}, as a function of the applied magnetic field.} 
\label{fig-hall}
\end{figure}
\begin{figure}[t]
\includegraphics[width=\columnwidth,clip]{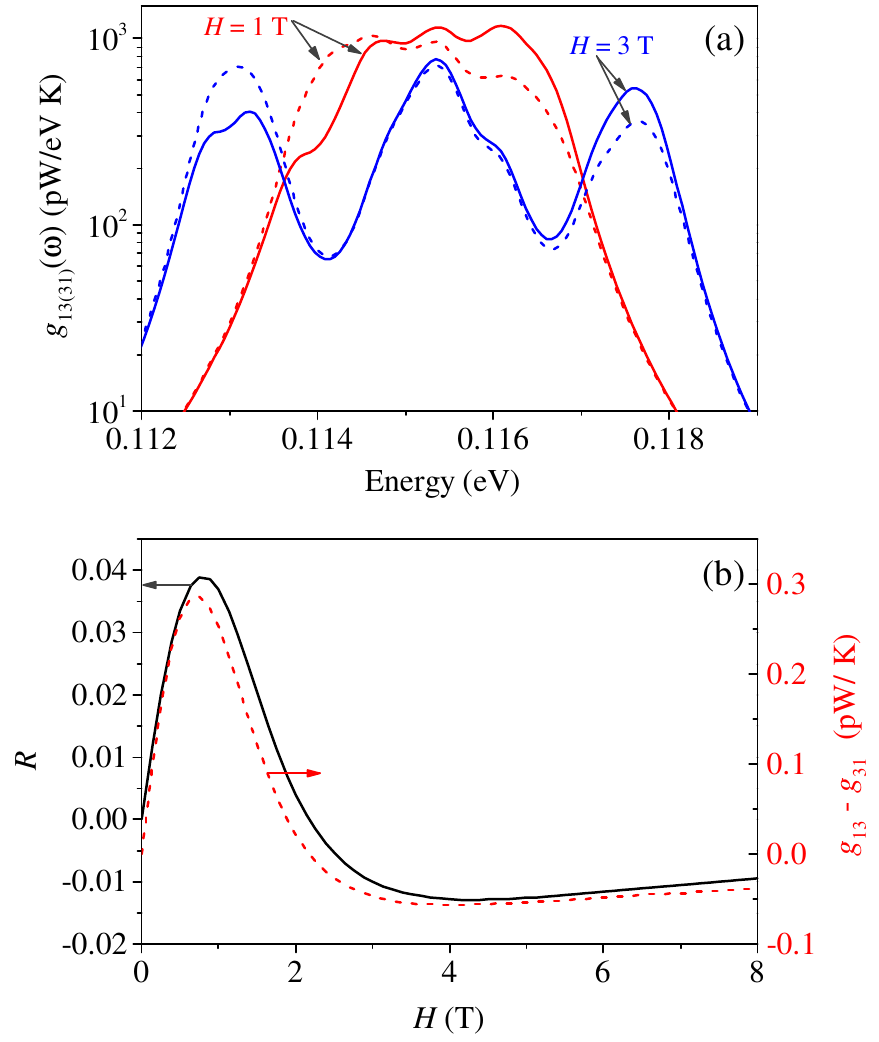}
\caption{Same as in Fig.~\ref{fig-hall}, but for a plasma frequency $\omega_p = 1.86 \times 10^{14}$ rad/s, which corresponds 
to a higher doping level, $n=1.36\times 10^{19}$ cm$^{-3}$, of the InSb.} 
\label{fig-hall-2}
\end{figure}

Now we can proceed to discuss the results of the photon thermal Hall effect that we obtain with our method. In Fig.~\ref{fig-hall}(a) 
we present the spectral thermal conductance, in both directions, between bodies 1 and 3 at thermal equilibrium (see 
Fig.~\ref{fig-system}). These results show that the magnetic field induces a directional spectral conductance as a result of 
the effect of the induced optical anisotropy under the collective surface modes of the system. This effect, explained in 
Ref.~\cite{Zhu2016} with the use of coupled mode theory, leads to both the existence of a heat persistent current at thermal 
equilibrium, characterized by the difference $g_{13}-g_{31}$, and the appearance of the photon thermal Hall effect, as is shown 
Fig.~\ref{fig-hall}(b). As we pointed out in preceding section, the difference between our analytical expression of 
Eq.~\eqref{trans-1} and that reported in Ref.~\cite{Ben-Abdallah2016} for point-like particles leads very to different results for 
the Hall coefficient (see Fig.~\ref{fig-hall}(b)). More precisely, we find that the occurrence of the photon thermal Hall signal requires 
more moderated intensities of the applied magnetic field than those reported in the original work. However, these field intensities are 
still relatively high due to the fact that the plasmonic resonances of the particles, which are required for the directional 
collective modes in the system, are located in the region of the electromagnetic spectra where the InSb polaritonic band 
$[\omega_T,\omega_L]$ is located. In order to enhance the sensitivity of the plasmonic resonances to the applied magnetic field, 
one can increase the carrier concentration in the InSb, and therefore the plasma frequency, as was done in Ref.~\cite{Guo2019}. 
Using a new plasma frequency of $\omega_p =  1.86 \times10^{14}$ rad/s, one obtains a higher sensitivity of the spectral conductance 
between bodies 1 and 3, as can be seen in Fig.~\ref{fig-hall-2}(a), and consequently, the persistent current and the photon 
thermal Hall effect occur for more moderate values of the magnetic field.

\begin{figure}[t]
\includegraphics[width=\columnwidth,clip]{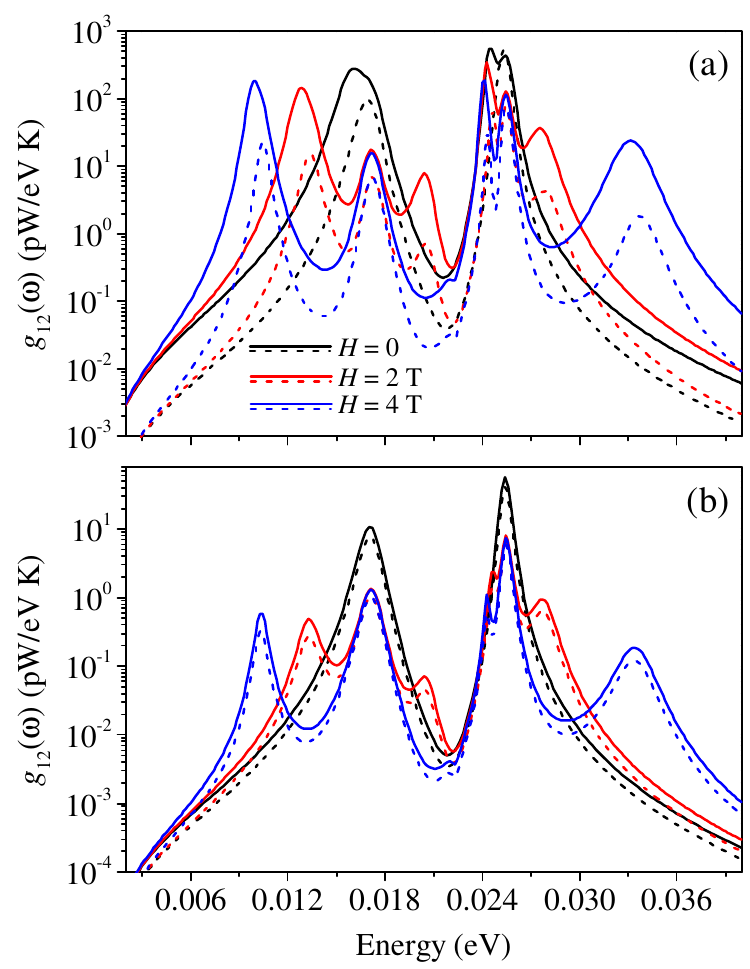}
\caption{Comparison between the arbitrary body approach Eq.~\eqref{trans-1} with $N_d=389$ dipoles (continuous lines) and the result
corresponding to a single point dipole (dashed lines), for different values of the static magnetic field. In both panels, spherical 
particles are assumed with a radius of 100 nm. Panels (a) and (b) corresponds to gap values $d=20$ nm and $d=120$ nm, respectively.}
\label{fig-point}
\end{figure}

Finally, we use our method to test the validity of the point-like approximation frequently used to analyze these many-body effects. 
For this purpose, in Fig.~\ref{fig-point} we compare the results for the spectral thermal conductance between spheres 1 and 2 
obtained with the exact approach developed in this work (solid lines) and those obtained assuming that the particles are point-like
(dashed lines). In this example all particles have the same radius of $r = 100$ nm and the vacuum gap is equal to $d=20$ nm for panel (a)
and $d=120$ nm for panel (b). As expected, when $d < r$ as in panel (a), the point-like approximation completely fails to reproduce
the exact results and it only becomes an acceptable approximation when $d > r$ (and the size of the particle is much smaller than 
the thermal wavelength). However, in such a limit, effects such as the persistent current or the photon thermal Hall effect turn out
to have very small amplitudes and they would be hard to measure in practice. This discussion illustrates the importance of using an exact
 method like the one presented in this work.

\section{Conclusions} \label{conclusions}

Motivated by the recent interest in the NFRHT in many-body systems featuring nonreciprocal objects, in this work 
we have extended the TDDA approach to deal with any number of optically anisotropic bodies of arbitrary size and
shape. We have illustrated this approach with the analysis of the photon thermal Hall effect and the appearance of a 
persistent current in a system comprising four MO spherical particles beyond the standard point-dipole approximation.
In the case of the photon thermal Hall effect we have amended the original analysis put forward in Ref.~\cite{Ben-Abdallah2016},
which was actually incorrect even in the limit of point-like particles. Moreover, our analysis not only shows that the
Hall effect survives when finite particle are used, but it also shows that it does not requires magnetic fields as 
high as originally reported. On the other hand, as already discussed in Ref.~\cite{Guo2019}, we have shown that the
existence of the photon thermal Hall effect in our many-body system is naturally accompanied by the appearance of a 
persistent directional heat current that does not violate the second law of thermodynamics.

Our TDDA-based approach can be used for a great variety of problems and it can also be straightforwardly extended to deal
with many different situations and phenomena. For instance, it is simple to adapt it to explore the non-additivity of the 
thermal emission of many-body systems or to study the possibility of controlling such an emission with external magnetic
fields (in the case of MO objects). On the other hand, since the TDDA is a volumen-integral-equation method, our 
approach is also well suited to deal with more complex situations in which there are nontrivial temperature profiles across 
the objects or when those objects are made of a combination of materials (giving rise to space-dependent permitivitties). For all 
those reasons, we think that the theoretical approach develop in this work is very valuable for the community of thermal
radiation and photonics in general. 

\acknowledgments

J.C.C.\ acknowledges funding from the Spanish Ministry of Science and Innovation (PID2020-114880GB-I00).

\end{document}